\documentclass[aps,prl,twocolumn,showpacs,superscriptaddress,groupedaddress]{revtex4-1}

\PassOptionsToPackage{caption=false}{subfig}
\usepackage{hyperref}
\usepackage[utf8]{inputenc}
\usepackage{amsmath}
\usepackage{mathtools}
\usepackage{graphicx}
\usepackage{svg}
\usepackage[euler]{textgreek}

\newcommand{\I}{\mathrm{i}}

\begin{document}

\title{Rheology of mixed motor ensembles}
\author{Justin Grewe} \author{Ulrich S. Schwarz}\email{schwarz@thphys.uni-heidelberg.de}
\affiliation{Institute for Theoretical Physics and Bioquant, Heidelberg University, Heidelberg, Germany}

\begin{abstract}
The rheology of biological cells is not only determined by their cytoskeletal networks, but
also by the molecular motors that crosslink and contract them. 
Recently it has been found that the assemblies of myosin II molecular motors
in non-muscle cells are mixtures of fast and slow motor variants. Using computer simulations
and an analytical mean field theory of a crossbridge model for myosin II motors, here we show that
such motor ensembles effectively behave as active Maxwell elements. We calculate
storage and loss moduli as a function of the model parameters
and show that the rheological properties cross over from viscous to elastic
as one increases the ratio of slow to fast motors. This suggests that
cells tune their mechanical properties by regulating the composition of their 
myosin assemblies.
\end{abstract}

\maketitle

The rheology of animal cells is essential for many physiological functions, including
the function of epithelial and endothelial cell layers under continous loading, e.g.\ in lung, skin,
intestines or vasculature \cite{trepat_mesoscale_2018}. It is also essential for single cell processes such as
cell migration and division, which are characterised by intracellular flows and deformations \cite{Kollmannsberger2011}. 
For these reasons, single cells and cell ensembles have been widely studied
using rheological approaches as commonly applied in materials science
\cite{Fabry2001,Micoulet2005,Balland2006,Fernandez2008,Fischer-Friedrich2016}.
Cells typically show a wide relaxation spectrum indicating the relevance of different time scales. Often power-law relaxation
spectra have been reported \cite{Fabry2001,Balland2006},
but there is also evidence for an upper cut-off at a maximum relaxation time \cite{Fischer-Friedrich2016}.
Despite this complexity of cell rheology, however, for many purposes linear viscoelasticity has
turned out to be a surprisingly good description of the effective mechanical properties of cells and cell monolayers
\cite{Mayer2010,Serra-Picamal2012,Vincent2015,Oakes2017,Saha2016,Vishwakarma2018,Wyatt2020}.

Cells actively control their mechanical properties mainly by changing the assembly status and activity of their actomyosin cytoskeleton.
Although much is known about the effective rheology of these networks
from the viewpoint of polymer physics \cite{Mizuno2007,Broedersz2014},
it is not clear how the microscopic properties of the different types of myosin motors contribute to cell rheology.
Recently it has been found that non-muscle cells co-assemble fast and slow isoforms of myosin II \cite{Beach2014, Beach2015}.
A very recent computational study showed that the electrostatic interactions between the
coiled-coils of the different isoforms leads to a rich energy landscape for mixed assembly and
can explain some aspects of their cellular localization \cite{kaufmann_electrostatic_2020}. 
While the fast myosin II isoform A is mainly found at the front of the cell, where
fast assembly and flow is required, the slow myosin II isoform B is
incorporated towards the back, where strong and long-lasting forces are required \cite{Beach2014, Beach2015}.

Here we explore the intriguing possibility that cells control their rheology by differential
assembly of their myosin minifilaments. We address this important question theoretically
by using a microscopic crossbridge model for small ensembles of myosin motors \cite{duke_molecular_1999,Erdmann2012,Erdmann2013,hilbert_kinetics_2013}, 
which earlier has been applied only to ensembles of one isoform \cite{Stam2015,Erdmann2016}. By extending this framework
to mixed ensembles and calculating their complex modulus, we show that such assemblies operate as active Maxwell elements
that can tune their rheology from viscous to elastic by increasing the ratio of slow versus fast motors.

\begin{figure}[tb]
\includegraphics[scale=1]{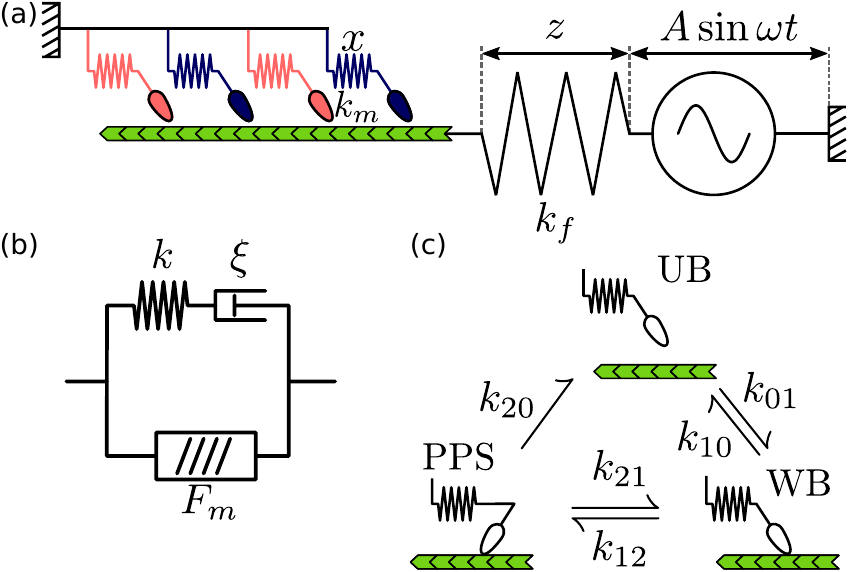}
\caption{Model. (a) Scheme used for rheology simulations of myosin II ensembles. 
The central spring has extension $z$ and spring constant $k_f$.
$A \sin \omega t$ is the oscillatory perturbation. 
The myosin crossbridges have motor strains $x$ and linker stiffness $k_m$. 
Blue and red myosin crossbridges denote the fast and slow isoforms A and B, respectively.
(b) Maxwell element with spring constant $k$ and friction coefficient $\xi$
in parallel with active motor force $F_m$. 
(c) Crossbridge model showing the three mechanochemical states 
(UB = unbound, WB = weakly bound, PPS = post-powerstroke) and the transition rates between them.}
\label{fig1}
\end{figure}

Fig.~\ref{fig1}a shows a schematic representation of the situation that we analyse here. 
A central spring, which could represent an optical trap or an elastic substrate, has extension 
$z$ and spring constant $k_f$. It is pulled from
two sides. On the right hand side, we have a mechanical motor that pulls
with fixed frequency $\omega$ and amplitude $A$. On the left hand side, 
we have a small ensemble of $N$ myosin II motor heads that walk towards the
barbed end of an actin filament. Each motor has a strain $x$ and a spring constant $k_m$. 
For the myosin II minifilaments in non-muscle cells, 
we typically would have $N=15$ \cite{Billington2013a,kaufmann_electrostatic_2020,Grewe2020}. 
From these $N$ motor heads, $N_a$ are assumed
to be of the fast isoform A. Then $N_b = N-N_a$ are of the slow isoform B.

We first show that in this setup, both the total mechanical system 
as well as the motor ensemble itself should behave effectively like an active Maxwell element 
as depicted in Fig.~\ref{fig1}b. The Maxwell element is the simplest possible viscoelastic model and
features a spring with spring constant $k$ and a dashpot with
friction coefficient $\xi$ in series; in an active Maxwell model, there is a constant
motor force $F_m$ operating in parallel. We assume that the motor ensemble
depicted in Fig.~\ref{fig1}a has a well-defined force-velocity relation $v(F)$,
with a free velocity $v_0$ at $F=0$ and vanishing velocity at the stall force $F=F_s$.
With all motors having the same crossbridge spring constant $k_m$, the motor ensemble
has an effective ensemble spring constant $k_e = n k_m$, with $n$ being the typical number of 
bound motors. Force balance leads to a differential equation for the extension $z$
\begin{align}
\dot{z} =  \frac{k_e}{k_e + k_f} \left[ v(k_f z) +A \omega \cos (\omega t) \right]\ .
\end{align}
After expanding the force-velocity relation around the stall force $F_s$ with a slope $v'(F_s) = - 1/\xi$, 
we easily can solve this equation:
\begin{equation}
z(t) =\frac{F_s}{k_f} +  C \exp \left(- \frac{k_t}{\xi} t \right) \\
+\frac{\omega k_t A }{k_f \sqrt{\omega^2 + (k_t/ \xi)^2}} \sin (\omega t - \delta)
\label{eq:osc}
\end{equation}
with the total spring constant $k_t = k_e k_f / (k_e + k_f)$ and $\tan \delta = k_t / \omega \xi$.
We obtain three terms, each with a clear physical meaning. The first term is the constant pull of the active Maxwell element,
thus $F_m = F_s$ as for a linear force-velocity relation \cite{Besser2007}. 
The second term is initial relaxation with a constant $C$ determined by the initial conditions.
The third term is our most important result: the system response is oscillatory
with the same frequency as the external perturbation, but with a loss angle $\delta$ 
that depends on the parameters of the motor ensemble. $\delta$ increases
from $0$ to $\pi/2$ with decreasing frequency. It also increases with 
increasing stiffness $k_t$ and decreases with increasing friction $\xi$. 
The calculated oscillations correspond to a complex modulus
\begin{align}
G^* = \frac{k_t \omega^2 \xi^2 }{k_t^2 +\omega^2 \xi^2} +  \I \frac{k_t^2  \omega \xi}{k_t^2 + \omega^2 \xi^2} \label{eq:maxwellmodulus}
\end{align}
which is exactly the result for a Maxwell model with the effective spring constant $k = k_t$
and a friction coefficient $\xi$. This result applies to the system that includes the 
external stiffness $k_f$. If we restrict ourselves to the motor ensemble, we also
obtain an active Maxwell model, with the same friction coefficient $\xi$ and
motor force $F_m$, but with the spring constant $k = k_e$ rather than $k_t$. This is equivalent to assuming an infinitely stiff environment, as $k_t \rightarrow k_e$ as $k_f \rightarrow \infty$.

In order to validate our prediction that motor ensembles should effectively behave as active
Maxwell systems, we conducted computer simulations of a microscopic 
crossbridge model for myosin II as shown in Fig.~\ref{fig1}c.
In our model, each of the $N$ crossbridges of the ensemble is in one of three mechanochemical states that are connected 
by force dependent transition rates \cite{Erdmann2012,Erdmann2013,Erdmann2016}. The transition from the unbound (UB) to the weakly bound (WB) state
occurs with rate $k_{01} =0.2\,$s$^{-1}$. The strain-dependent reverse rate is $k_{10}(x) = k_{10}^{0} \exp(k_m x^+/f_s)$, 
where $k_{10}^{0}=0.004\,$s$^{-1}$ is the rate at zero strain, $k_m=0.3\,$pN/nm is the crossbridge stiffness, $x^+ = \mathrm{max}(0,x)$ is the positive part of the crossbridge strain $x$ and $f_s=10.55\,$pN/nm the internal force scale. 
From the WB-state, the motor can perform a powerstroke of distance $d = 8$ nm,
which also determines how far the motors can advance. The rate for transition into the
post-powerstroke (PPS) states is governed by  the difference in elastic energy stored 
in the crossbridges, $\Delta E_\mathrm{el}$, and the free energy of ATP-hydrolysis, $\Delta G=-60\,$pN$\,$nm. 
We use $k_{12/21}= k_\mathrm{ps} \exp ( \pm \beta (\Delta E_\mathrm{el} + \Delta G)/2)$ 
with $k_\mathrm{ps}=1000\,$s$^{-1}$. Finally, the unbinding from the PPS-state is modeled as a \textit{catch-slip} bond, 
i.e. $k_{20}(x) = k_{20}^{0 a/b} (\Delta_c \exp (-k_m x^+/f_c) + \Delta_s \exp (k_m x^+/f_s))$. Here $\Delta_c=0.92$ is the 
fraction following the \textit{catch-path} at zero force with force scale $f_c=1.66\,$pN, $\Delta_s = 0.08$ 
is the fraction following the \textit{slip-path} at zero force. For the faster NM IIA motors 
we use $k_{20}=1.71\,$s$^{-1}$, while for the slower NM IIB we use $k_{20}=0.35\,$s$^{-1}$ \cite{Erdmann2016}. 

\begin{figure}[tb]
\includegraphics[scale=1]{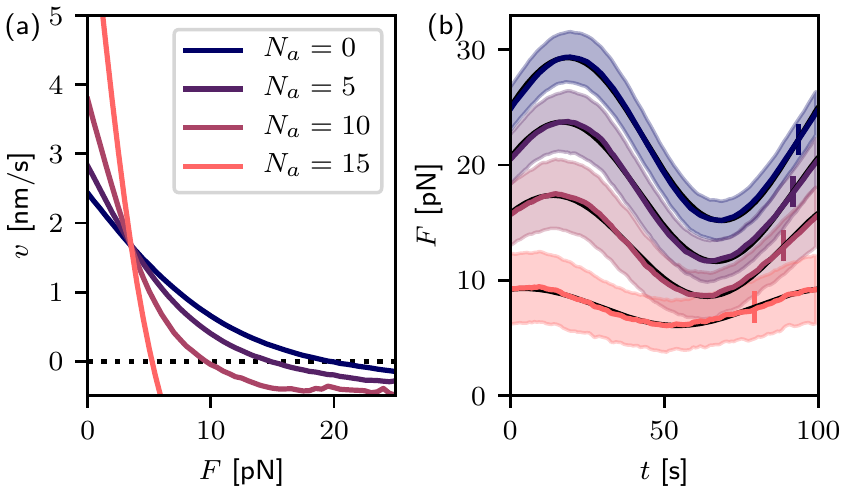}
\caption{Computer simulations of the crossbridge model for mixed ensembles with $N=15$ motor heads. 
(a) Force-velocity relations. Zero crossings define the stall force $F_s$ and the linear
slope around this point the friction coefficient $\xi$. Throughout the paper we color code the fraction 
of NM IIA motors as a gradient from dark blue for  $N_a=0$ to light red for $N_a = N$. (b) Force
on central spring as a function of time. The colored areas denote the region of one standard deviation.
The solid lines denote fits with sine functions and the vertical tics indicate where the new period starts.}
\label{fig2}
\end{figure}

By mixing $N_a$ and $N_b$ motors, we now can explore how the minifilament composition determines
its effective rheology. Fig.~\ref{fig2}a shows that indeed our microscopic model leads to a well-defined force-velocity relation
for each value of $N_a$, which then defines both the stall force $F_s$ and the effective friction coefficient $\xi$. We find that
as the number of fast motors $N_a$ is increased, both $F_s$ and $\xi$ decrease. Fig.~\ref{fig2}b
shows that as predicted by Eq.~\eqref{eq:osc}, the system response is oscillatory and can be fit well with a sine wave $F(t) = F_s +  \Delta F \sin (\omega t-\delta)$ 
with an amplitude $\Delta F$, loss angle $\delta$ and force offset $F_s$ that depend on model parameters. Here we have identified the constant offset force with the stall force as suggested by Eq.~\eqref{eq:osc} for $A=C=0$.
We see that the loss angle $\delta$ increases while the 
amplitude and the constant offset force $F_s$ decrease 
with increasing NM IIA content. This suggests that the system crosses over from 
viscous to elastic as the fast myosin IIA motors are replaced by the slow myosin IIB motors.

In order to achieve a deeper
understanding and an accurate mapping between the microscopic motor rates and the effective
Maxwell rheology, we next developed a self-consistent mean field treatment of our crossbridge
model for motor ensembles \cite{Erdmann2012,Erdmann2013}. In steady state binding and unbinding from the track is balanced.
Assuming the powerstroke is performed immediately after binding and approximating the stall force of the ensemble as the sum of the single motor stall forces, we have $F_s = n k_m d$ and $n = k_{01} N / (k_{20}(d)+ k_{01})$.
Using the formerly derived relation between speed and the number of bound
motors when all of them are in the PPS-state \cite{Erdmann2013}, 
$v(F) = (N-n)k_{01} (d-F/(n k_m))/((n+1))$,
we can derive the total spring constant $k_t$ and the friction coefficient $\xi$
as functions of the mechanochemical rates, the environmental stiffness $k_f$ and the ensemble size $N$:
\begin{align}
k_t&=\frac{k_m N
    }{ 1+\frac{ k_{20}(d)}{k_{01}}+\frac{k_m N}{k_f}}\ , \label{eq:keff}\\
\xi&=  \frac{k_m n(n+1)}{(N-n)k_{01}} \label{eq:etaeff}.
\end{align}
To approximate these quantities for motor ensembles with heterogeneous composition, 
we reason that the relevant factor is the time spent by each motor bound to the filaments. 
For the rates, this implies that we should take the harmonic mean
\begin{equation}
k_{20}^0(N_a, N_b) = \frac{N_a + N_b}{N_a/k_{20}^{0a}+ N_a/k_{20}^{0b}},
\label{eq:rates}
\end{equation}
with $N_a,\, k_{20}^{0a},\, N_b,\, k_{20}^{0b}$ the total number and the off-rates of NM IIA and B heads, respectively.

\begin{figure}[t]
\includegraphics[scale=1]{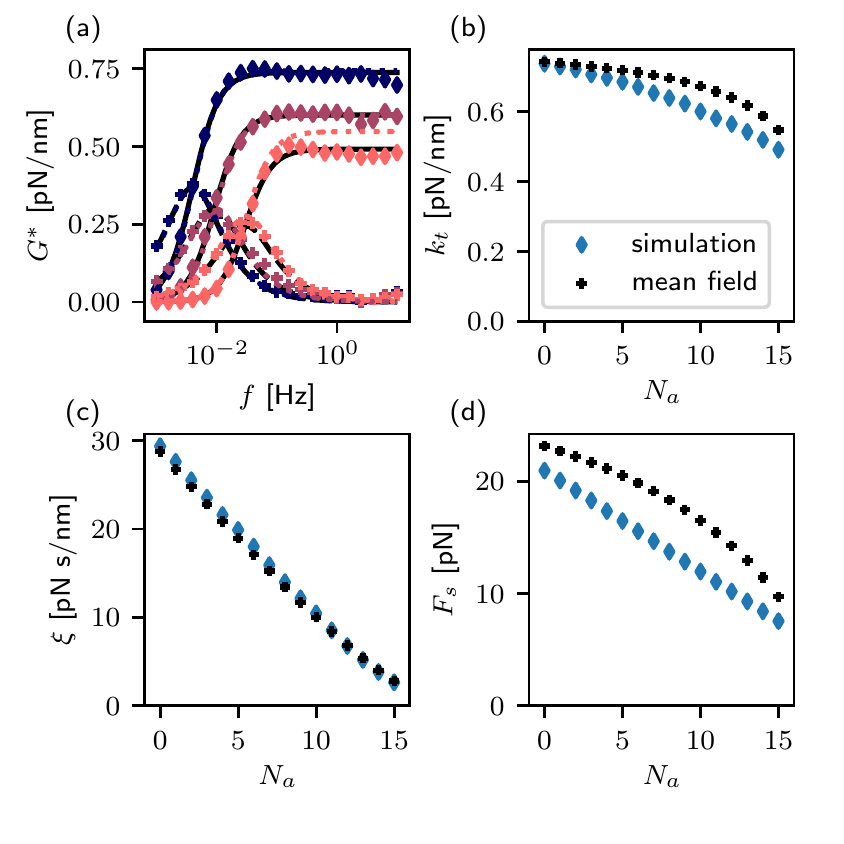}
\caption{Mechanical response as a function of ensemble isoform content for ensemble size $N=15$ 
and external stiffness $k_f=1\,$pN/nm. (a) The dynamic modulus $G^*$ of a ensembles with $N_a \in \{0, 10, 15 \}$ (dark blue, light violet, light red). The diamonds and crosses denote the storage and loss as determined from the simulation, respectively, while the black lines are Maxwell model fits. The interrupted lines are the result of the analytical mean field theory. (b, c, d) Spring constant $k_t$, friction coefficient $\xi$ and stall force $F_s$ as a function of NM IIA content $N_a$. Blue diamonds and black crosses denote the results obtained from the simulation and the analytical mean field theory, respectively.}
\label{fig3}
\end{figure}

In Fig.~\ref{fig3}(a) we show that the results from the computer simulations (symbols, obtained
via calculating $G^* = \Delta F (\cos \delta + \I \sin \delta)/A$) and from the analytical mean field 
theory (interrupted lines, obtained by combining Eqs.~\eqref{eq:maxwellmodulus}-\eqref{eq:rates}) 
are in very good agreement with each other. In addition we plot fits of
Eq.~\eqref{eq:maxwellmodulus} to the results of the computer
simulations (solid lines). The mean field theory does not perform perfectly
for $N_a = N$ because in this case of only fast motors, it can happen that all motors are unbound at the same time,
while the mean field theory assumes that $n$ always has a finite value.
In Figs.~\ref{fig3}(b) - (d) we show that the mean field theory also performs well for predicting 
total spring constant $k_t$, friction coefficient
$\xi$ and stall force $F_s$ as a function of the number of fast motors. For (b) and (c), we obtained these
values through fits of Eq.~\eqref{eq:maxwellmodulus} to the results of the computer simulations. 
One sees that the effective spring constant decreases slightly with increasing NM IIA content,
which is consistent with the lower duty ratio of a single NM IIA motor compared to NM IIB. The friction coefficient decreases markedly with increasing NM IIA starting from $\xi\approx 30\,$pN$\,$s/nm without NM IIA and ending at $\xi \approx 2\,$pN$\,$s/nm with purely NM IIA for a motor ensemble with 15 motors. The stall force in (d) has been obtained directly from the computer simulation as
the force at the beginning of a new period. One sees that 
the stall force $F_s$ goes from $\sim 20\,$pN/nm for ensembles of pure NM IIB ensembles to $\sim 10\,$pN/nm 
for pure NM IIA ensembles.

\begin{figure}[t]
\includegraphics[scale=1]{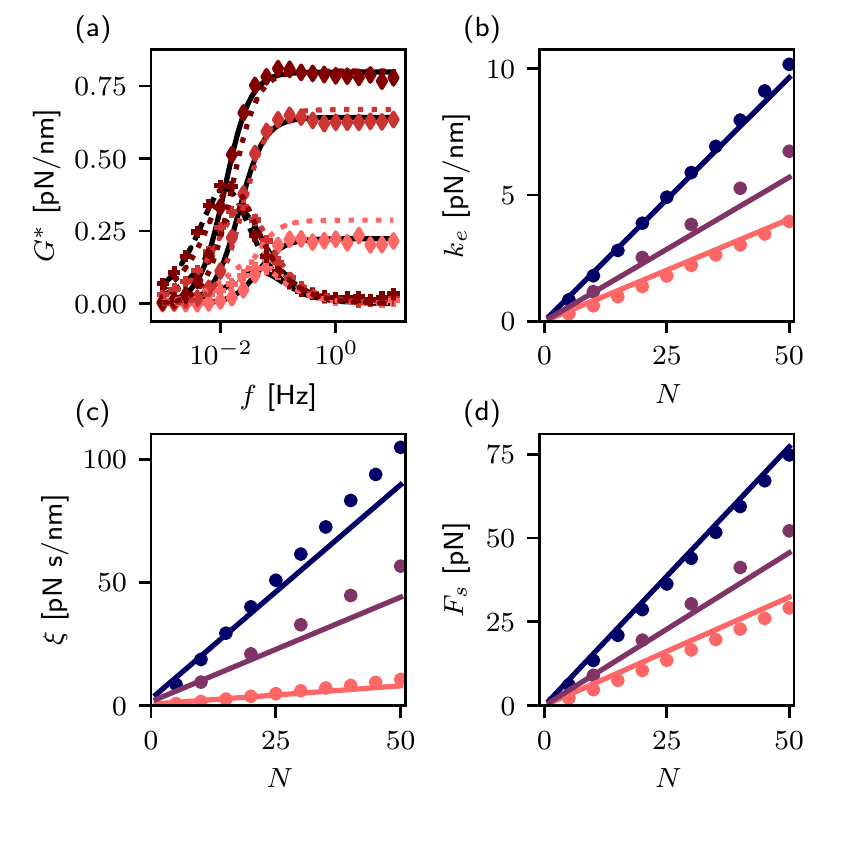}
\caption{Mechanical response as a function of ensemble size. (a) Dynamic modulus for NM IIA ensembles of size $N = N_a \in \{5, 25, 50\}$ (light to dark red)  with $k_f=1\,$pN/nm (for symbol description see Fig.~\ref{fig3}(a)). (b, c, d) motor ensemble spring constant $k_e$, friction coefficient $\xi$ and ensemble stall force $F_s$ as a function of ensemble size $N$ for $N_a \in \{ 0, N/2, N\}$ (dark blue, violet, light red). Symbols and lines represent the computer simulations and the analytical mean field theory, respectively.}
\label{fig4}
\end{figure}

Non-muscle myosin II assemblies are very dynamic and often change their size
in the cellular context. We therefore next investigated the 
size-dependence of the mechanical response, as shown in Fig.~\ref{fig4}. 
Ensemble spring constant $k_e=n k_m$, friction coefficient $\xi$ and stall force $F_s$
rise linearly with size. This suggests, that one can think of a single motor of the ensemble as an active Maxwell element which in the context of the ensemble is in parallel to others. The linear relationships also motivate calculating the effective Young's modulus $E=k_e l / \pi r^2 $, viscosity $\eta = \xi l / \pi r^2$ and the active stress $\sigma_m = F_s / \pi r^2$ generated by one half-minifilament of length $l\approx 150\,$nm with a typical crosssectional radius of $r \approx 20\,$nm, i.e.\ the distance the heads typically splay outward from the center of the filament with $N=15$ motors \cite{Billington2013a}. We find $E = 160 - 460\,$kPa, $\eta = 0.45 - 4.8\,\mathrm{MPa}\, \mathrm{s}$ and $\sigma_a =2 - 5 \,$kPa for pure NM IIA and NM IIB ensembles, respectively.
While the values for $E$ and $\eta$ are higher than typical cellular values because they
describe only the condensed situation in the myosin assemblies, the active
stress $\sigma_a$ is exactly the order of magnitude measured e.g.\ with
monolayer stress microscopy \cite{Trepat2009}.

The viscoelastic relaxation time of the ensemble follows as $\tau = \xi / k_e = \eta / E \approx 10\, s$, which is
exactly the order of magnitude observed in laser cutting experiments \cite{Colombelli2009,Kassianidou2017}.
Our results suggest that the exact relaxation time should depend on the mix of 
NM IIA and B motors in the stress fiber, as indeed reported experimentally \cite{Tanner2010, Lee2018}. 
In particular, the relaxation time of mature stress fibers can be reduced by suppressing NM IIB gene expression,
as predicted here \cite{Chang2015}.

\begin{figure}[t]
\includegraphics[scale=1]{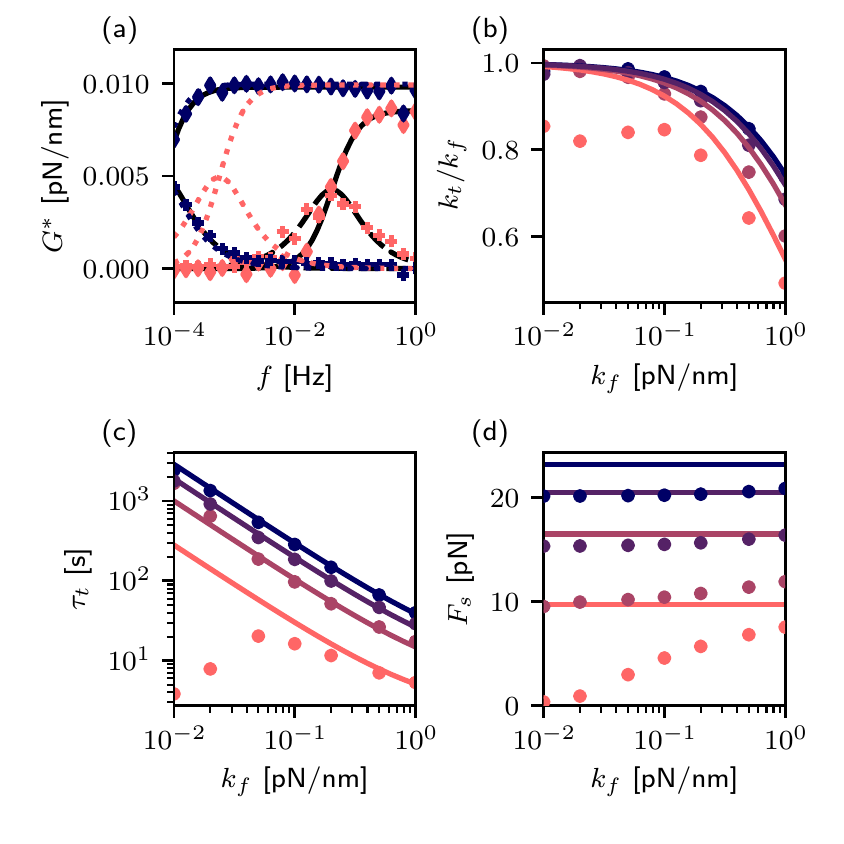}
\caption{Mechanical response as a function of external stiffness $k_f$. 
(a) Dynamic modulus for NM IIA and NM IIB ensembles of size $N=15$ at 
$k_f=0.01\,$pN/nm (dark blue, light red) (for symbol description see Fig.~\ref{fig3}(a)). 
(b, c, d) Normalized stiffness $k_t/k_f$, total relaxation time $\tau_t$ and stall force $F_s$ 
as a function of external stiffness $k_f$ for $N_a \in \{0, 5, 10, 15\}$ 
(dark blue, dark violet, light violet, light red). Symbols and lines represent the 
simulations and the analytical mean field theory, respectively.}
\label{fig5}
\end{figure}

Cells respond very sensitively to the stiffness of their environment and we therefore
also investigated the role of the external stiffness $k_f$ as shown in Fig.~\ref{fig5}. 
For sufficiently high external stiffness, the normalized stiffness $k_t/k_f$ and the total 
relaxation time $\tau_t=\xi/k_t$ both 
decrease with increasing environment stiffness, while the stall force $F_s$ 
does not depend much on the external stiffness. 
Interestingly, the mean field approximation fails dramatically for pure NM IIA  
ensembles of size $N=15$ for environments less stiff than $k_f=0.1\,$pN/nm,
because in this regime often all motors unbind simultaneously. Thus
minifilaments containing mainly the fast isoform A cannot build
up forces in a soft environment; this however becomes possible when
external stiffness becomes larger than internal stiffness. For isoform B,
this effect is less pronounced. These results agree with the biological
expectation that rigidity sensing by migrating cells has to occur at the
front, where myosin IIA is localized.

In summary, here we have proposed a scale-bridging theory that relates 
the microscopic stochastic dynamics of the motor crossbridges to a
macroscopic linear viscoelastic model, namely the active Maxwell model.
Our main result is that incorporating more NM IIB motors makes
the system more elastic, in agreement with their physiological role to stabilize
the rear end of migrating cells. Our results also agree with the 
role of NM IIB motors to increase the relaxation time of stress fibers
as revealed by laser cutting. Finally we found that myosin IIA can
better sense external stiffness and might do so at the front
of migrating cells. Overall, our work shows that cells can 
modulate the interaction with their mechanical
environment not only through their cytoskeletal networks, but also through
the size and composition of their motor ensembles.

\begin{acknowledgments}
We thank Jordan Beach, Tom Kaufmann, Kai Weißenbruch, Martin Bastmeyer and 
Elena Kassianidou for helpful discussions and Falko Ziebert for critical reading of the manuscript.
This work is supported by the Deutsche Forschungsgemeinschaft (DFG, German Research Foundation) under Germany's Excellence Strategy EXC 2181/1 - 390900948 (the Heidelberg STRUCTURES Excellence Cluster).
\end{acknowledgments}


%

\end{document}